\documentstyle[12pt,aasms4]{article}

\slugcomment{}

\begin{document}

\title{Halo-Stripping in Galactic Collisions}

\author{P. J. S. Watson }
\affil{Physics Department
Carleton University,
Ottawa 
Canada K1S 5B6}

\begin{abstract}
 Galactic collisions are normally modeled in a CDM model by assuming the DM consists of a small number of very massive objects. This note shows that the behaviour of a CDM halo during collisions depends critically on the mass of the particles that make it up, and in particular, all halo particles below a certain characteristic mass are likely to be lost.
\end{abstract}

\keywords{galaxies: dark matter, merging galaxies}

\section{Introduction}

	There is now a sizeable literature on collisions between galaxies, and computer models show a remarkable resemblance to observations of interacting galaxies (see, e.g. \cite{Bar92}) .  Typically these involve the numerical solution of N-body collisionless Boltzmann equations, where $N \sim 10^{4}$. This implies that the effective mass of a dark-matter (DM) body is $10^6 M_\odot$ and one may wonder whether the results are independent of this assumption. In particular, it would be interesting to know whether the behaviour of cold dark matter (CDM) halo particles in a collision is effectively the same whether they are massive neutrinos or MACHOs. 

	In this note we point out that CDM haloes may behave in radically different ways depending on the mass of the particles. Specifically haloes composed of particles lighter than a certain mass $m_0$ ($m_0 \sim10^{16} $kg) will be stripped off a galaxy during a collision, while heavier ones will remain with the galaxy and would presumably become part of a merged galaxy. We derive a specialized form of the Chandrasekhar dynamical friction formula which is the gravitational analogue of the Bethe-Bloch formula, and then apply this to a highly idealized galaxy collision. Finally we offer some speculations on how this will affect galaxy populations.

\section{Dynamical Friction}

	Consider a particle of mass $\mu$, which is incident on a gravitationally interacting gas consisting of particles of mass m and number density n. The dynamical  friction formula (\cite{Cha43, Bin87} can be written
\begin{equation}
{{d\vec v_m} \over {dt}}=-2\pi \ln \left( {1+\Lambda ^2} \right)G^2\mu \left( {m+\mu } \right)\int {f\left( v \right)}{{\left( {\vec v-\vec v_m} \right)} \over {\left| {\vec v-\vec v_m} \right|^3}}d^3v
\end{equation}

 We take the gas to be at rest, and the incident particle to be moving with a velocity v and simplify to the case $m = \mu$. None of these assumptions are critical to the argument. It is then easy to transform (1) to
\begin{equation}
{{dE} \over {dx}}=-{{8\pi G^2\mu ^2\rho } \over {v^2}}\ln \left( \Lambda  \right)
\end{equation}

where $ \rho$ is the mass density of the gas. The argument of the log is very large, so the log can be replaced by a dimensionless constant $C_0$. The resulting equation can be derived from the Bethe-Bloch (see, e.g. \cite{Ros52}) equation by a judicious change of variables. Note the occurrence of the $\mu^2$ term, which is crucial to the following argument: qualitatively one can see that the cross-section is proportional to $\mu$ and the mean momentum transfer is also proportional to $\mu$. It is also somewhat counter-intuitive that the force increases as $1 \over {v^2}$ at low velocities.
	To see how this affects a galactic halo, consider the collision process shown below \placefigure{fig1}. The galaxies are assumed to have a spherically symmetric halo of particles of mass $\mu$ and density $\rho(r)$, and to be moving with a relative speed of $v_\infty$. It is then trivial to integrate (2) to give
\begin{equation}
v^4=v_\infty ^4-16\pi C_0G^2\mu \int {\rho (r)ds^{}}
\end{equation}

where the integral is along the path followed by the particle.
	To proceed further we need to have a definite model for the density distribution in the galaxy. The model must give flat rotation curves out beyond the edge of the visible galaxy, and yet have a finite mass. For simplicity we choose
\begin{equation}
M(r)=M_\infty \left( {1-e^{-\left( {r\over r_0} \right)}} \right)
\end{equation}

which gives a density function
\begin{equation}
\rho \left( r \right)=\rho _0^{}{{e^{-y}} \over {y^2}},y={r \over {r_0}}
\end{equation}

(This is, of course, not a good approximation for the core of a galaxy, but fortunately this is not relevant to the following argument). If the rotational speed of the galaxy is $v_0$ we get
\begin{equation}
M_\infty ={{r_0v_0^2} \over G},\rho _0={{M_\infty } \over {4\pi r_0^3}}
\end{equation}
This allows (4) to be integrated giving
\begin{equation}
v^4=v_\infty ^4-32\pi G^2\mu r_0C_0I\left( {b_0} \right)_{},I\left( {b_0} \right)=\int\limits_{b_0}^\infty  {{{e^{-y}} \over {y\sqrt {y^2-b_0^2}}}}dy\approx {{e^{-b_0}} \over {b_0}}
\end{equation}

where $b_0 = {b \over {r_0}}$ is the scaled impact parameter and the approximation for the integral is good to about $50\%$ for the range $.01<b_0<2$. By inspection of  one can see that a particle will only be stopped in a galactic  collision if its mass was larger than the characteristic mass $\mu_0$  given by
\begin{equation}
\mu _0={{v_\infty } \over {2\left( {2\pi G^2\rho _0C_0I\left( {b_0} \right)} \right)^{1 \over 4}}}
\end{equation}
\section{An idealized Galaxy Collision}.
	To give some idea of the value of $\mu_0$, we can take plausible values of $r_0 = 50$ 
kPc and the orbital velocity $v_0 = 200 km s^{-1}$ and further assumes that the galaxies have a
velocity $v_\infty = 100 km s^{-1} $ when they collide. This gives rise to $M_\infty = 10^{42}  kg \sim  10^{12} M_\odot$ , $\rho _0 ~ 10^{-22} kg m^{-3}$ and $C_0 \sim 300$ (for $\mu = 10^{-27}$  kg) to 30 ($\mu = 10^{30}$ kg). This gives rise to a characteristic mass $\mu _0 \sim 10^{16}$ kg. Loosely speaking, any particle with a mass of less than  $\mu_0$  would not be
 stopped in the collision. Note that this argument does not apply to gas or dust, which are affected
 by conventional friction. Hence we may expect that baryonic matter would behave in the way 
predicted by the N-body simulations, and  in particular MACHOÕs would not be lost to the merged 
galaxy. Possibly asteroids which are sufficiently massive that the conventional frictional force 
on them would be relatively small and which were not gravitationally bound to a solar system 
would also be lost.

	However, particles which interact only via gravitational interactions only, which means 
neutrinos or WIMPs (we use WIMPs generically in what follows) would behave very 
differently. WIMP haloes would be stripped off during a collision of this kind, with a time scale 
of  ${r_0 \over {v_\infty} } \sim10^8$ yrs.  To see what would happen to the stars in such a 
collision, consider a collision between two identical spiral galaxies, which occurs in such a way 
that the cores coalesce immediately.  Assuming that the luminous matter obeys a similar mass 
distribution function

\begin{equation}
\rho \left( r \right)=\rho _0^{}r_{0^{}}^2{{e^{-{r \over {r_1}}}} \over {r^2}}
\end{equation}	
					
so that the galaxy is entirely luminous matter at the centre, and $r_1 \approx 10$ kPc is the radius of the luminous
 matter. In this case, stars at the centre of the galaxies would be essentially 
unaffected by the loss of the halo, stars out to a critical radius $R \sim 1.59 r_1$ would be 
forced into increasingly elliptical orbits and any stars beyond R would no longer be bound.

	The most serious weakness in this model is the assumption that the ÒtargetÓ gas is at 
rest. Since obviously nothing is known about the velocity of WIMP haloes, it is difficult to 
correct this. One might argue that the WIMPs form a gas with a BB temperature of (say) 2 K,
 which would correspond to velocities in the range of a few $ms^{-1}$, in which case the 
assumption is obviously valid. Alternatively the  WIMP halo could co-rotate with the central 
galaxy, or (more plausibly) form a gravitationally self-bound system, in 
which case the virial theorem gives

\begin{equation}
\left\langle {v^2} \right\rangle ^{1over{ 2}}=\left( {{{GM_\infty \ln (2)} \over {r_0}}} \right)^{1over{2}}
\end{equation}				
The latter two assumptions both imply speeds of ~ 200 $km s^{-1}$ whichare comparable to the  collision speeds. This would require much more detailed modeling than is possible here, but one 
would expect that some fraction of the halo with a small velocity in the final rest system of the 
merged galaxy would be retained, while the majority would be  stripped. 	

The idea that haloes 
are stripped in collisions is in qualitative agreement with two results. Firstly \cite{Dub96} 
 show that tidal tails form during galaxy mergers appear to be considerably longer than would 
be predicted by N-body simulations. They note that the best estimates of dark matter derived 
from  tidal tails in merging galaxies such as NGC 4038/39 and NGC 7252 imply a value about 
1/3 of that derived from spiral  galaxies. This could be understood if much of the halo material 
is lost after the collision. If, as is generally believed, elliptic galaxies are formed by collisions 
of spirals, they should be almost devoid of haloes. Galaxies in a cluster would have lost much of 
their DM haloes during repeated collisions, and the DM would be bound to the cluster as a whole.
  This is, of course, consistent with observations of (e.g.) the Coma cluster. 

	One could regard this as strong circumstantial evidence that the main component of haloes are 
WIMPs. Obviously if this can be confirmed by more detailed models, it would be of considerable 
importance.

\section{Conclusion} This calculation suggests that the behaviour of the DM haloes during a galactic collision depends critically on the mass of the particles that make it up. If the DM objects are relatively light (both neutrinos and WIMPs are light in this context), then the halo will be lost totally, while heavy haloes  would be retained as predicted by the N-body simulations.   In other words, this process provides a way to differentiate between a particle type solution and a MACHO type solution to the DM problem.

\acknowledgments

	This work was suggested by a talk given by Jayanne English, and I am grateful to her for a useful correspondence. Conversations with Bob Carnegie and Steve Godfrey are acknowledged. This work was partly supported by NSERC.



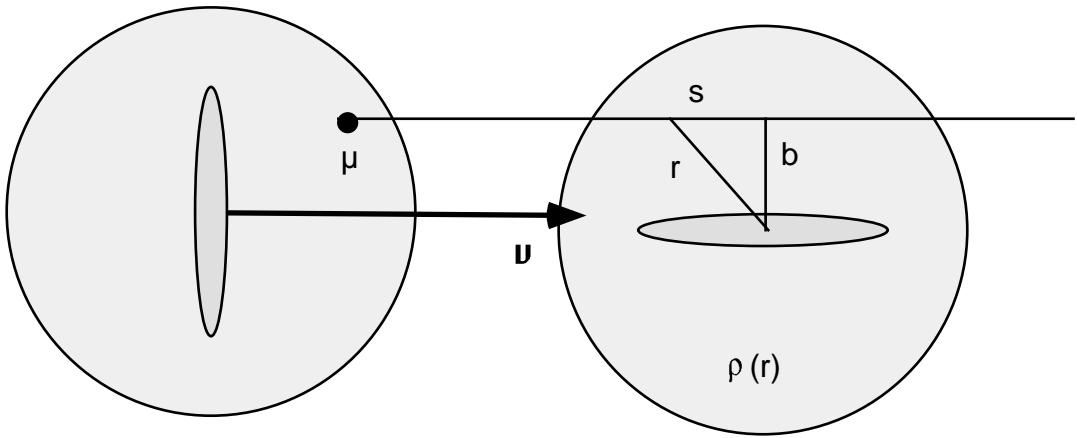
\end{document}